\documentclass{aastex}          
\usepackage{spr-astr-addons}    
\usepackage{epsfig}
\usepackage{graphicx}
\usepackage{color}

\begin{document}
%
\title{Non-linear damping of visco-resistive Alfv\'{e}n waves in solar spicules}

\author{Z.~Fazel, and H.~Ebadi\altaffilmark{1}}
\affil{Astrophysics Department, Physics Faculty,
University of Tabriz, Tabriz, Iran\\
e-mail: \textcolor{blue}{z$_{-}$fazel@tabrizu.ac.ir}}

\altaffiltext{1}{Research Institute for Astronomy and Astrophysics of Maragha,
Maragha 55134-441, Iran.}

\begin{abstract}
Interaction of Alfv\'{e}n waves with plasma inhomogeneities generates phase mixing which can lead to dissipate Alfv\'{e}n waves and
to heat the solar plasma. Here we study the dissipation of Alfv\'{e}n waves by phase mixing due to viscosity and resistivity variations with height.
We also consider nonlinear magnetohydrodynamic (MHD) equations in our theoretical model.
Non-linear terms of MHD equations include perturbed velocity, magnetic field, and density.
To investigate the damping of Alfv\'{e}n waves in a stratified atmosphere of solar spicules,
we solve the non-linear MHD equations in the $x-z$ plane.
Our simulations show that the damping is enhanced due to viscosity and resistivity gradients.
Moreover, energy variations is influenced due to nonlinear terms in MHD equations.

\end{abstract}


\section{Introduction}
\label{sec:intro}
Alfv\'{e}n wave phase mixing has been studied extensively as a possible mechanism for coronal heating \citep{Heyvaerts1983, Parker1991, Hood1997a, Nakara1997,
De99, Tsik2002, Mocan2008, Mcla2011}. When the medium has a density gradient perpendicular to the magnetic field, the Alfv\'{e}n speed is a function of the
transverse coordinate.
The generation of transverse gradients in the wave leads to a strong increase in the dissipation of Alfv\'{e}n wave energy due to viscosity and/or resistivity.
\citet{Heyvaerts1983} were first to suggest that phase mixing of Alfv\'{e}n waves in coronal plasmas could be a primary mechanism in heating of corona. Then
several authors have studied the phase mixing in different conditions and ways. \citet{Parker1991} investigated the effect of a density and/or a temperature
gradient in the direction of vibration of a transverse Alfv\'{e}n wave. A strong coupling of waves on different lines of force, producing a coordinated mode
that was not subject to simple phase mixing, was resulted. \citet{Hood1997a, Hood1997b} derived a self similar solution of Alfv\'{e}n wave phase mixing in both
open and closed magnetic topologies. \citet{Nakara1997} extended the model of \citet{Heyvaerts1983}, considering the non-linear excitation of fast magnetosonic
waves by phase mixing of Alfv\'{e}n waves in a cold plasma with a smooth inhomogeneity of density across a uniform magnetic field. They found that this
non-linear process could be a possible mechanism of indirect plasma heating by phase mixing through the excitation of fast waves. \citet{Botha2000} by considering
a developed stage of Alfv\'{e}n waves phase mixing showed that the non-linear generation of fast modes by Alfv\'{e}n waves has little effect on the classical phase mixing.
\citet{Tsik2002} considered the interaction of an impulsively-generated, weakly non-linear MHD pulse with
a 1D density inhomogeneity in the three-dimensional regime, in an ideal plasma. They found that phase mixing remains a relevant paradigm. \citet{Mcla2011}
have investigated the non-linear, non-ideal behavior of Alfv\'{e}n wave propagation and phase mixing over long timescales. They found that the equilibrium density profile
is significantly modified by both the flow of density due to visco-resistive heating and the non-linear response to the localized heating through phase mixing.

Spicules are one of the most fundamental components of the solar chromosphere. They are seen in spectral lines at the solar limb at speeds of
about $20-25$~km~s$^{-1}$ propagating from photosphere into the magnetized low atmosphere of the Sun \citep{Tem2009}. Their diameter and length varies from spicule
to spicule having the values from $400$~km to $1500$~km and from $5000$~km to $9000$~km, respectively.
The typical lifetime of them is $5-15$ min. The typical electron density at heights where the spicules are observed
is approximately $3.5\times10^{16}-2\times10^{17}$ m$^{-3}$, and their temperatures are estimated as $5000-8000$ K \citep{bec68, ster2000}.
\citet{Kukh2006} and \citet{Tem2007} observed their transverse oscillations with the estimated period of $20-55$ and $75-110$ s by analyzing the height
series of $H\alpha$ spectra in solar limb spicules observed. Recently, \citet{Ebadi2012a} based on \emph{Hinode}/SOT observations estimated the oscillation period
of spicule axis around $180$ s. They concluded that the energy flux stored in spicule axis oscillations is of order of coronal energy loss in quiet Sun.
\\In this paper we are interested to study the non-linear Alfv\'{e}n wave propagation and phase mixing in a stratified atmosphere, i.e., the spicule.
Section $2$ gives the basic equations and theoretical model. In section $3$ numerical results are presented and discussed, and a brief summary is followed in section $4$.

\section{Theoretical modeling}
\label{sec:theory}
\subsection{The equilibrium}
We consider effects of the stratification due to gravity in $2$D x-z plane in the presence of steady flow and shear field.
The phase mixing and the dissipation of propagating Alfv\'{e}n waves are studied in a region with nonuniform Alfv\'{e}n velocity both along
and across the spicule axis.
Non-ideal MHD equations in the plasma dynamics are as follows:
\begin{equation}
\label{eq:mass} \frac{\partial \mathbf{\rho}}{\partial t} + \nabla
\cdot (\rho \mathbf{v}) = 0,
\end{equation}
\begin{equation}
\label{eq:momentum} \rho\frac{\partial \mathbf{v}}{\partial t}+
\rho(\mathbf{v} \cdot \nabla)\mathbf{v} = -\nabla p + \rho
\mathbf{g}+ \frac{1}{\mu_{0}}(\nabla \times \mathbf{B})\times
\mathbf{B}+ \rho\nu(z)\nabla^2\mathbf{v},
\end{equation}
\begin{equation}
\label{eq:induction} \frac{\partial \mathbf{B}}{\partial t} = \nabla
\times(\mathbf{v} \times \mathbf{B})+ \eta(z)\nabla^2\mathbf{B},
\end{equation}
\begin{equation}
\label{eq:divergence} \nabla \cdot \mathbf{B} = 0,
\end{equation}
\begin{equation}
\label{eq:state} p = \frac{\rho RT}{\mu}.
\end{equation}
where $\mu_{0}$ is the vacuum permeability and $\mu$ is the mean molecular weight. $\nu(z)$ is the viscosity coefficient which is defined for a fully ionized and
collision-dominated $H$ plasma. It should be noted that the Coulomb logarithm has a weak dependence on temperature and density variations which is in agreement with the
presence of the transition region. It is given by
\begin{equation}
\label{eq:viscos} \rho\nu(z)= 2.2\times 10^{-17}T_{0}(z)^{5/2}~    kg m^{-1}s^{-1},
\end{equation}
and $\eta(z)$ is the resistivity coefficient which is defined as a typical value in the solar chromosphere and corona \citep{Prie1982} by
\begin{equation}
\label{eq:resist} \eta(z)= (8\times 10^{8}-10^{9})T_{0}(z)^{-3/2} ~  m^2s^{-1},
\end{equation}
the temperature profile in Eqs.~\ref{eq:viscos}, and~\ref{eq:resist} is taken as a smoothed step function, i.e.:
\begin{equation}
\label{eq:temp}
 T_{0}(z)= \frac{1}{2}T_{c}\left [1+d_{t}+(1-d_{t})\tanh(\frac{z-z_{t}}{z_{\omega}})\right],
\end{equation}
here, $d_{t}=T_{ch}/T_{c}$ which $T_{ch}$ is the chromospheric temperature at its lower part and $T_{c}$ denotes the coronal
temperature that is separated from the chromosphere by the transition region. $z_{w}=200$~km is the width of transition region which is located
at the $z_{t}=2000$~km above the solar surface. We put $T_{ch}=15\times10^{3}$~K and $T_{c}=1\times10^{6}$~K.

\subsection{The perturbations}
\begin{figure}
\centering
\includegraphics[width=8cm]{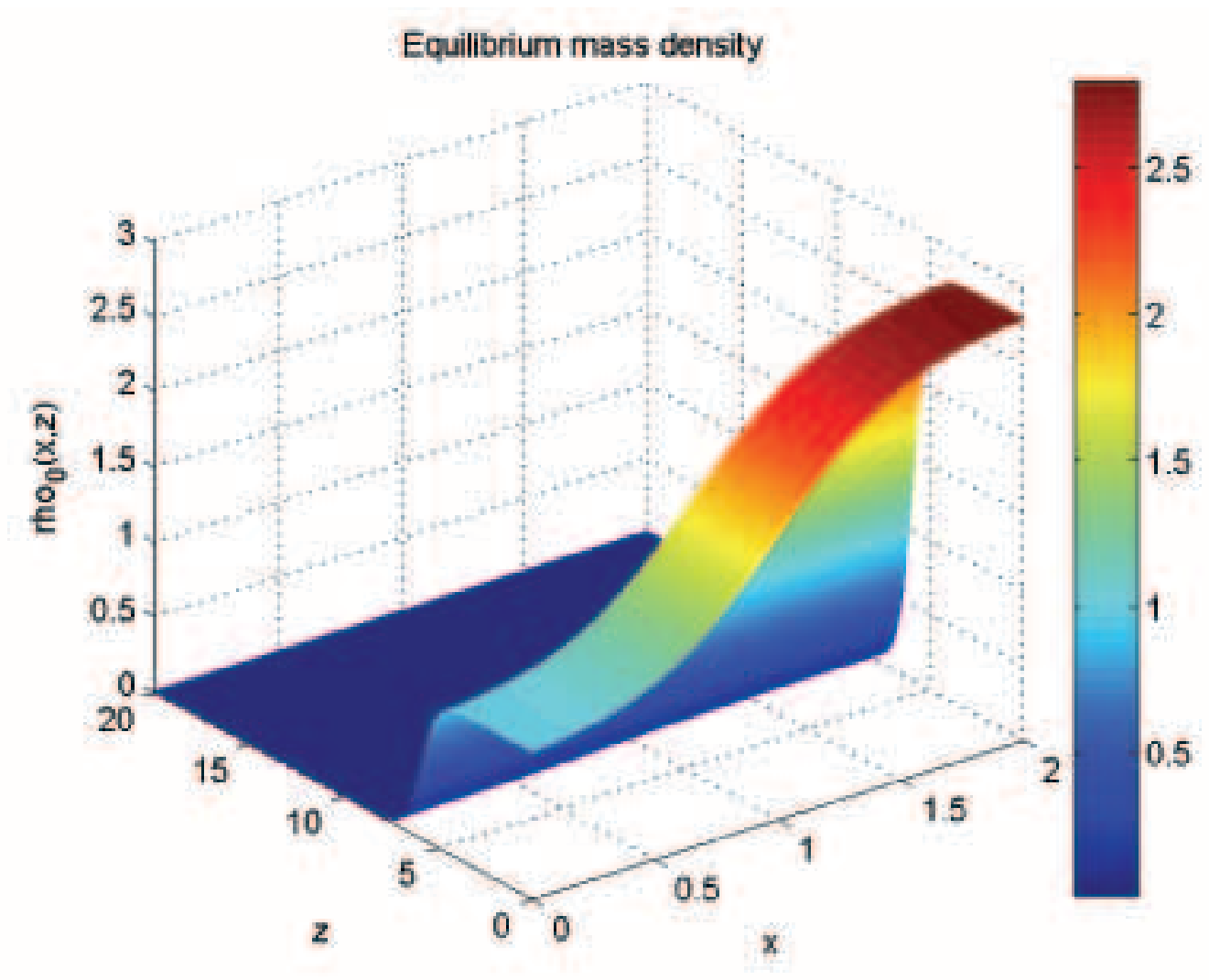}
\includegraphics[width=8cm]{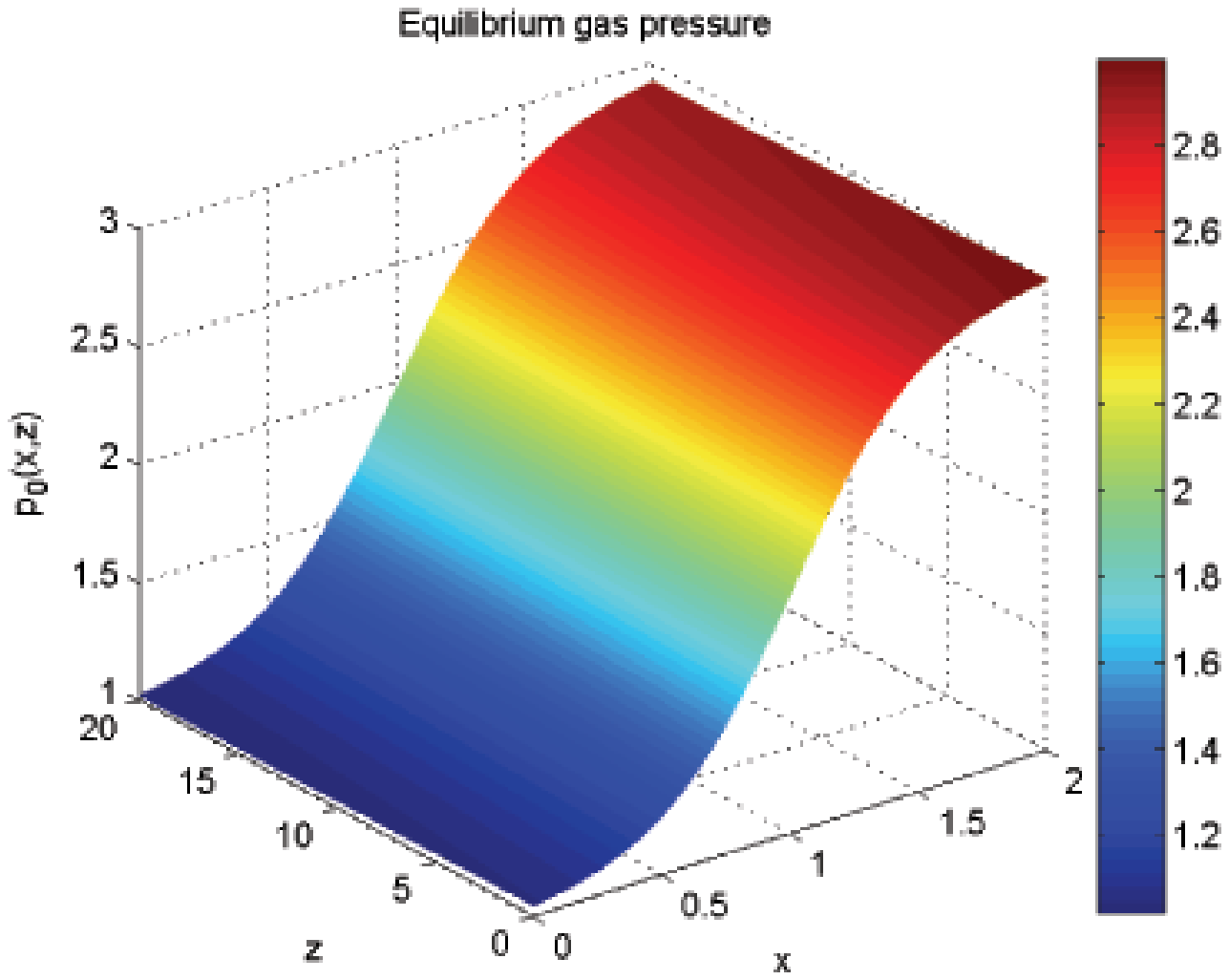}
\includegraphics[width=8cm]{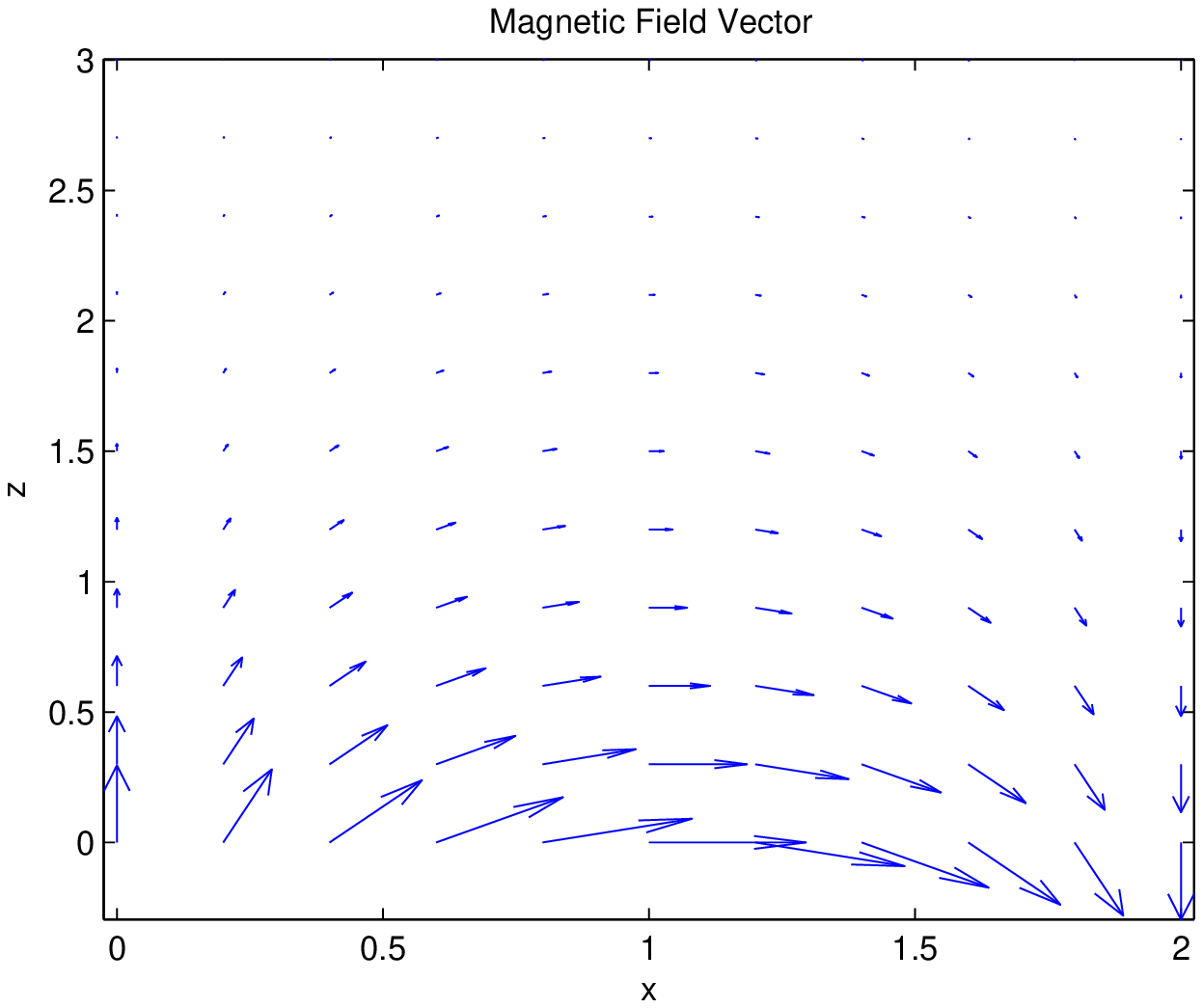}
\caption{The panels from top to bottom represent equilibrium mass density, gas pressure, and magnetic field lines, respectively. \label{fig1}}
\end{figure}

We assume that spicules are highly dynamic with speeds that are significant fractions of the Alfv\'{e}n speed. Perturbations are assumed to be independent of y, i.e.:
\begin{eqnarray}
\label{eq:perv}
  \textbf{v} &=& v_{0} \hat{k} + v_{y}(x,z,t) \hat{j}, \nonumber\\
  \textbf{B} &=& B_{0}e^{-k_{b}z} \left[\cos[k_{b}(x-a)]\hat{i}-\sin[k_{b}(x-a)]\hat{k} \right] \nonumber\\
  & & + b_{y}(x,z,t) \hat{j}
\end{eqnarray}
where $a$ is the spicule radius. The equilibrium sheared magnetic field is two-dimensional and divergence-free \citep{Del2005,Tem2010}.
\\ Since the equilibrium magnetic field is force-free, the pressure gradient is balanced by the gravity force, which is assumed
to be $\textbf{g}$=$-g\hat{k}$ via this equation:
\begin{equation}
\label{eq:balance}
 -\nabla p_{0}(x,z) + \rho_{0}(x,z) \textbf{g}=0,
\end{equation}
the pressure in an equilibrium state is:
\begin{equation}
\label{eq:presse}
 p_{0}(x,z)= p_{0}(x)~\exp\left(-\int^{z}_{z_{r}}\frac{dz'}{\Lambda(z')}\right).
\end{equation}
and the density profile is written in the following form:
\begin{equation}
\label{eq:density}
\rho_{0}(x,z)= \frac{\rho_{0}(x)T_{0}}{T_{0}(z)}~\exp\left(-\int^{z}_{z_{r}}\frac{dz'}{\Lambda(z')}\right),
\end{equation}
where $\rho_{0}(x)$ is obtained from the Alfv\'{e}n velocity for a phase mixed and stratified atmosphere due to gravity which is assumed to be \citep{De99, Karami2009}:
\begin{equation}
\label{eq:densityx}
 \rho_{0}(x)= \rho_{0} [2+ \tanh(\alpha(x-a))]^{-2},
\end{equation}
and
\begin{equation}
\label{eq:scale}
 \Lambda(z)= \frac{RT_{0}(z)}{\mu g},
\end{equation}
where $\rho_{0}$ is the plasma density at $z=5000$~km, $\alpha$ controls the size of inhomogeneity across the magnetic field.
In Figure~\ref{fig1} we present the equilibrium mass density, gas pressure, and magnetic field lines.

\begin{figure}
\centering
\includegraphics[width=7cm]{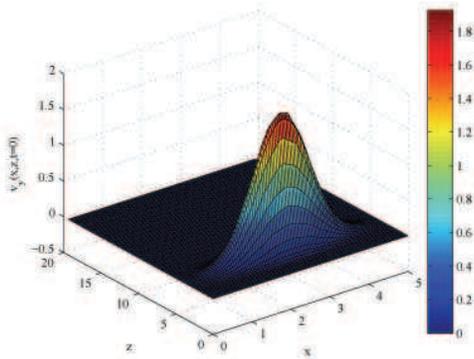}
\caption{The initial wave packet for $\omega =0.8a$ is showed.  \label{fig2}}
\end{figure}

The non-linear dimensionless MHD equations with these assumptions are:
\begin{eqnarray}
\label{eq:dens}
   \frac{\partial \rho_{1}(x,z)}{\partial t}+ v_{0}\frac{\partial \rho_{0}(x,z)}{\partial z}+ v_{0}\frac{\partial \rho_{1}(x,z)}{\partial z} &=& 0
\end{eqnarray}
\begin{eqnarray}
\label{eq:veloy}
   (\frac{\partial v_{y}}{\partial t}+ v_{0}\frac{\partial v_{y}}{\partial z}) &=& \nonumber
   \\
   \frac{\left[B_{0x}(x,z)\frac{\partial b_{y}}{\partial x}+B_{0z}(x,z)\frac{\partial b_{y}}{\partial z}\right]}{\rho_{0}(z)+\rho_{1}(x,z)}+ \nu(z)\nabla^{2}v_{y},
\end{eqnarray}
\begin{eqnarray}
\label{eq:mag}
  \frac{\partial b_{y}}{\partial t}+ v_{0}\frac{\partial b_{y}}{\partial z}  &=&  \nonumber
  \\ \left[B_{0x}(x,z)\frac{\partial v_{y}}{\partial x}+B_{0z}(x,z)\frac{\partial v_{y}}{\partial z}\right] + \eta(z)\nabla^{2}b_{y},
\end{eqnarray}
where densities, velocities, magnetic field, time and space coordinates are normalized to $\rho_{\rm 0}$ (the plasma density at dimensionless $z=6$), $V_{A0}$, $B_{\rm 0}$, $\tau$,
and $a$ (spicule radius), respectively. Also the gravity acceleration is normalized to $a^{2}/\tau$. Second terms in the left hand side of Eqs.~\ref{eq:veloy} and ~\ref{eq:mag} present the effect of steady flows. Multiplying of $\rho_{1}(x,z)$ in the first term of Eq.~\ref{eq:veloy} in the left hand side, gives us the non-linear contribution of MHD waves. Eqs.~\ref{eq:veloy}, and \ref{eq:mag} should be solved under following initial and boundary conditions:
\begin{eqnarray}
\label{eq:icv}
  v_{y}(x,z,t=0) &=& V_{A0}\exp \left[-\frac{(x-x_{0})^{2}+(z-z_{0})^{2}}{\omega^{2}}\right] \nonumber\\
  b_{y}(x,z,t=0) &=& 0,
\end{eqnarray}
where $\omega$ is the width of the gaussian packet.
\begin{eqnarray}
\label{eq:bc}
  v_{y}(x=0,z,t) = v_{y}(x=2,z,t) = 0, \nonumber\\
  b_{y}(x=0,z,t) = b_{y}(x=2,z,t) = 0, \nonumber\\
  \rho_{1}(x=0,z,t) = \rho_{1}(x=2,z,t) = 0.
\end{eqnarray}
Figure~\ref{fig2} shows the initial wave packet given by Eq.~\ref{eq:icv} for $\omega=0.8a$ ($a$ is the spicule radius).

\section{Numerical results and discussion}
To solve the coupled Eqs.~\ref{eq:veloy}, and~\ref{eq:mag} numerically,
the finite difference and the Fourth-Order Runge-Kutta methods are used to take the space and time derivatives, respectively.
The implemented numerical scheme is using by the forward finite difference method to take the first spatial derivatives with the truncation error of ($\Delta x$),
which is the spatial resolution in the $x$ direction. The order of approximation for the second spatial derivative in the finite difference method is $O((\Delta x)^2)$.
On the other hand, the Fourth-order Runge-Kutta method takes the time derivatives in the questions. The computational output data are given in $17$ decimal
digits of accuracy.
\\We set the number of mesh-grid points as~$256\times256$. In addition, the time step is chosen as $0.0005$, and the system length
in the $x$ and $z$ dimensions (simulation box sizes) are set to be ($0$,$5$) and ($0$,$20$).

The parameters in spicule environment are as follows:
$a=250$~km (spicule radius),
$\omega=0.8a=200$~km (the width of Gaussian packet), $L=5000$~km (Spicule length), $v_{0}=25$~km s$^{-1}$,
$n_{e}=11.5\times10^{16}$~m$^{-3}$, $B_{0}=1.2\times10^{-3}$~Tesla, $T_{0}=14~000$~K, $g=272$~m s$^{-2}$, $R=8300$~m$^{2}$s$^{-1}$k$^{-1}$
(universal gas constant), $V_{A0}=75$~km/s, $\mu=0.6$, $\tau=20$~s, $\rho_{0}=1.9\times10^{-10}$~kg m$^{-3}$, $\alpha =2$,
$p_{0}=3.7\times10^{-2}$~N m$^{-2}$, $\mu_{0}=4\pi \times10^{-7}$~Tesla m A$^{-1}$, $z_{r}=5000$~km (reference height), $z_{w}=200$~km,
$z_{t}=2000$~km, $x_{0}=1000$~km, $z_{0}=125$~km, $H= 750$~km, $\eta=10^{3}$~m$^2$ s$^{-1}$, and $k_{b}=\pi/2$ (dimensionless wave number normalized to $a$).

\begin{figure}
\centering
\includegraphics[width=8cm]{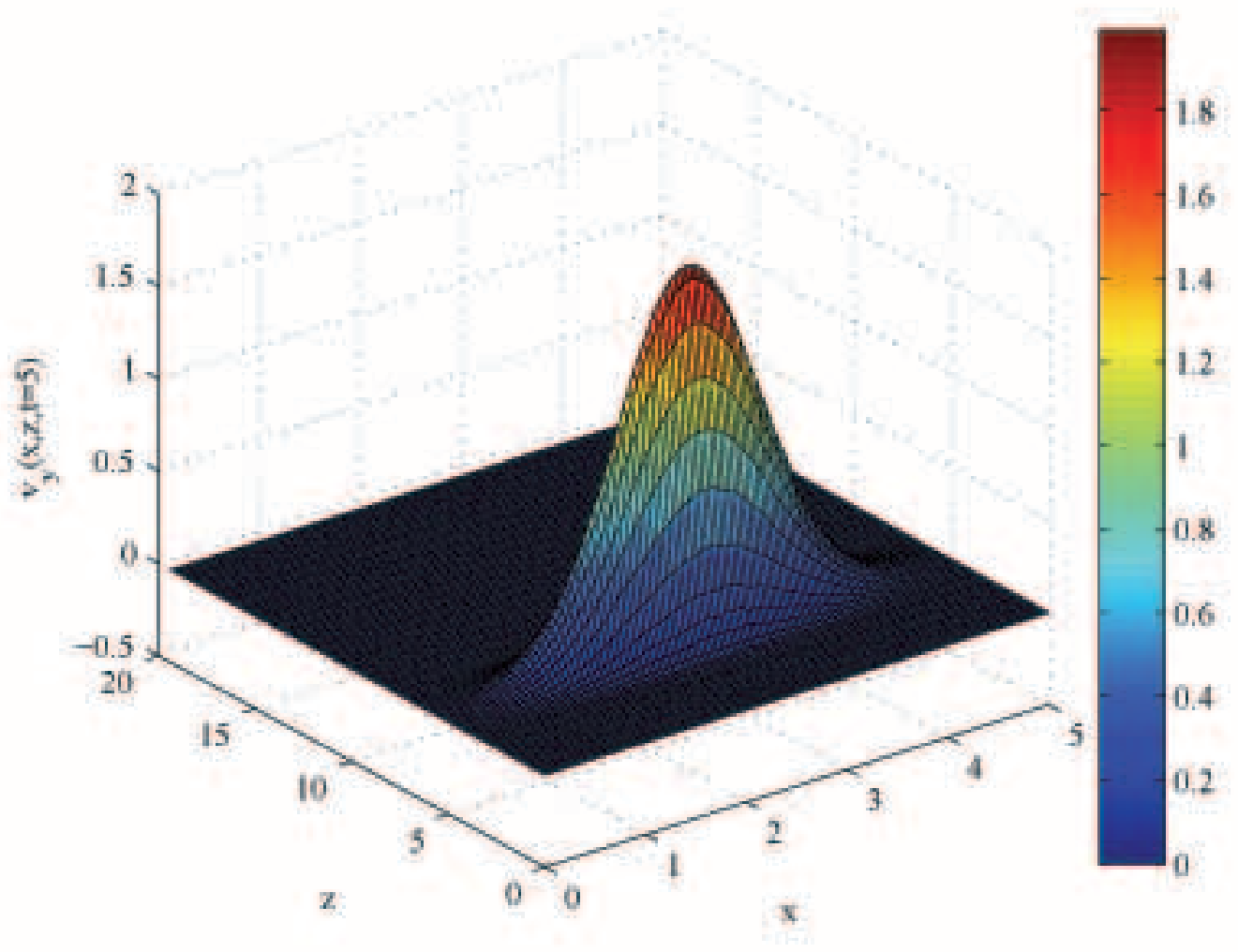}
\includegraphics[width=8cm]{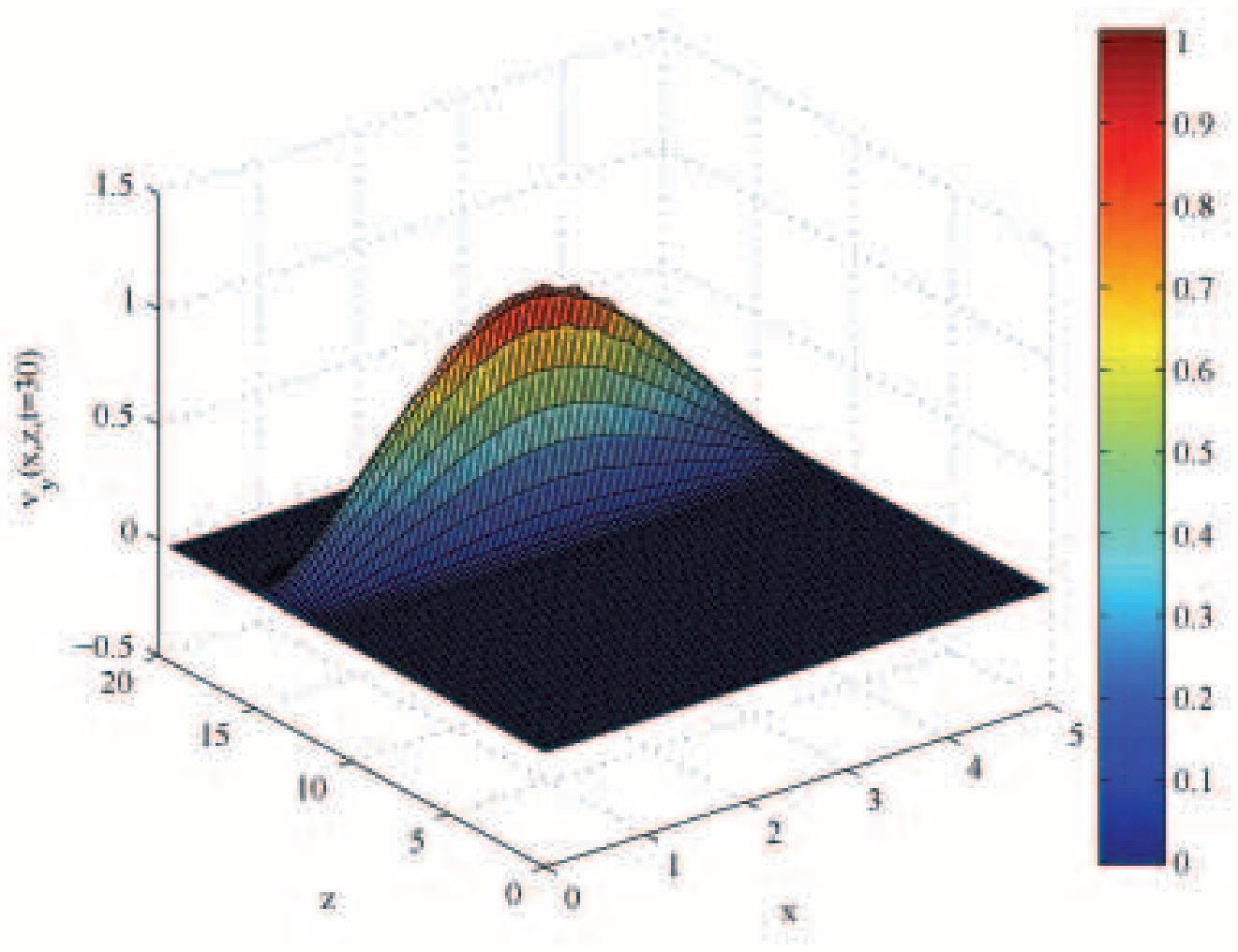}
\includegraphics[width=8cm]{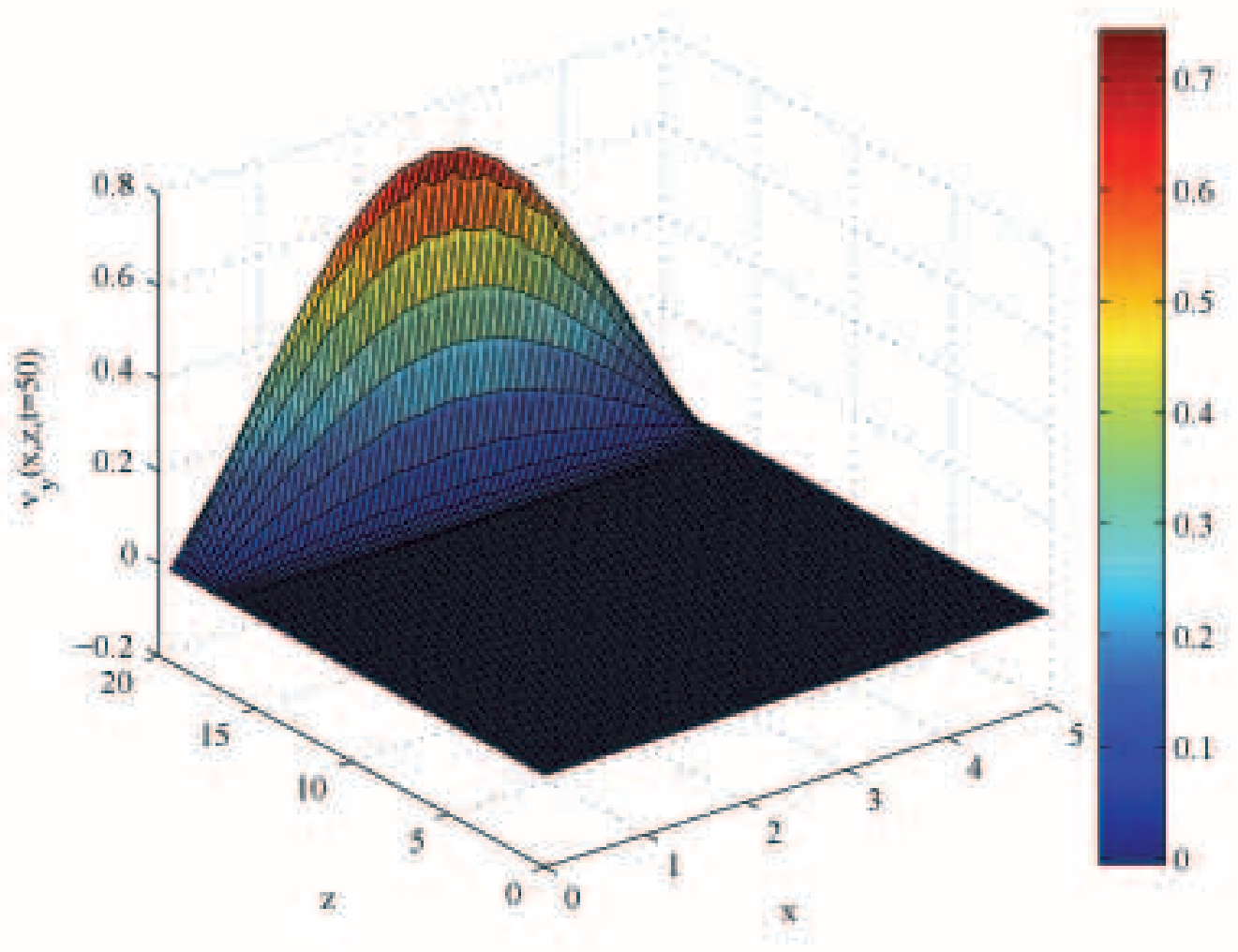}
\caption{The $3D$ plots of the transversal component of the perturbed velocity with respect to
$x$, $z$ in $t=5 \tau$~s, $t=30 \tau$~s, and $t=50 \tau$~s for $k_{b}=\pi/2$. \label{fig3}}
\end{figure}

\begin{figure}
\centering
\includegraphics[width=8cm]{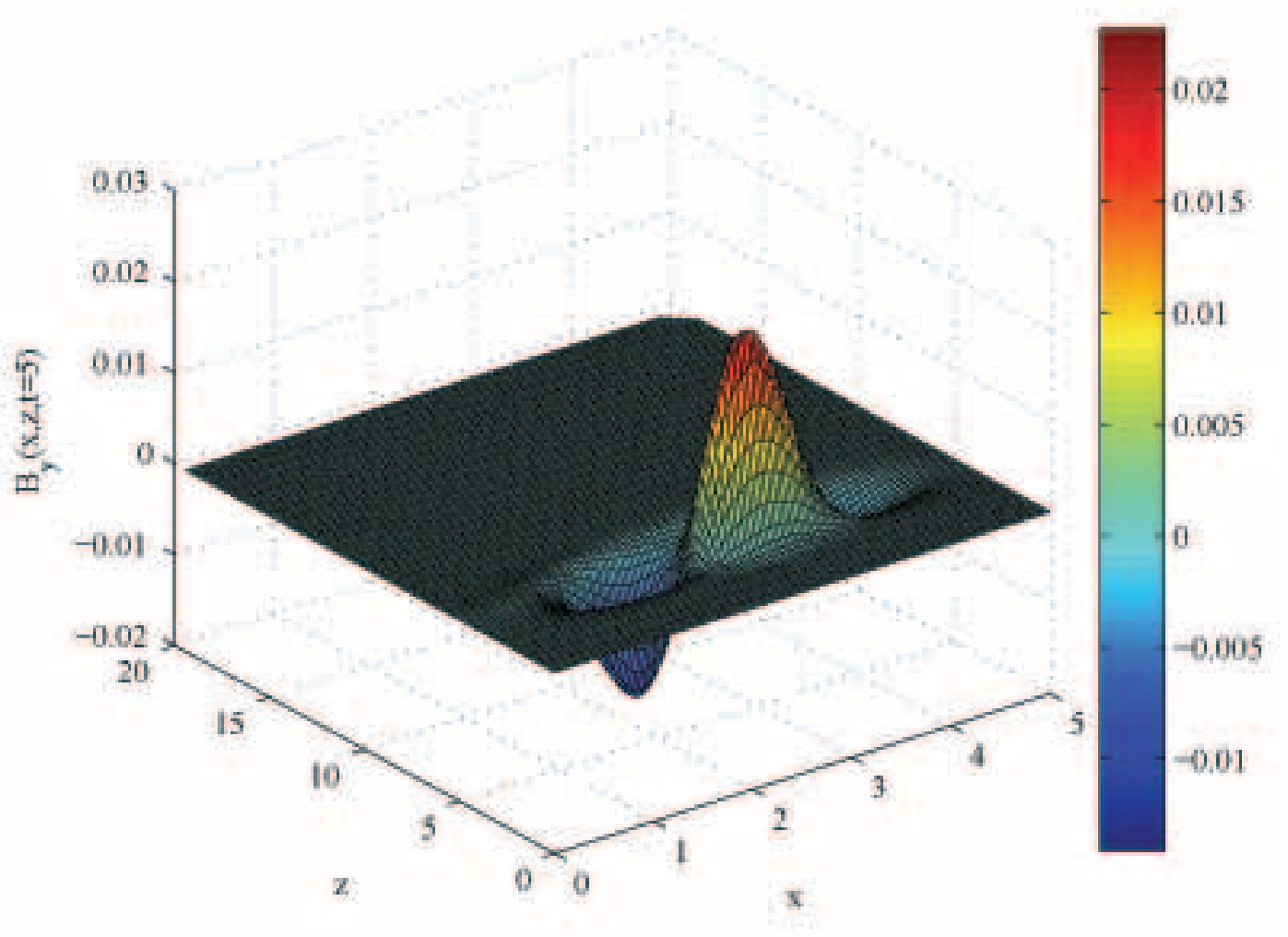}
\includegraphics[width=8cm]{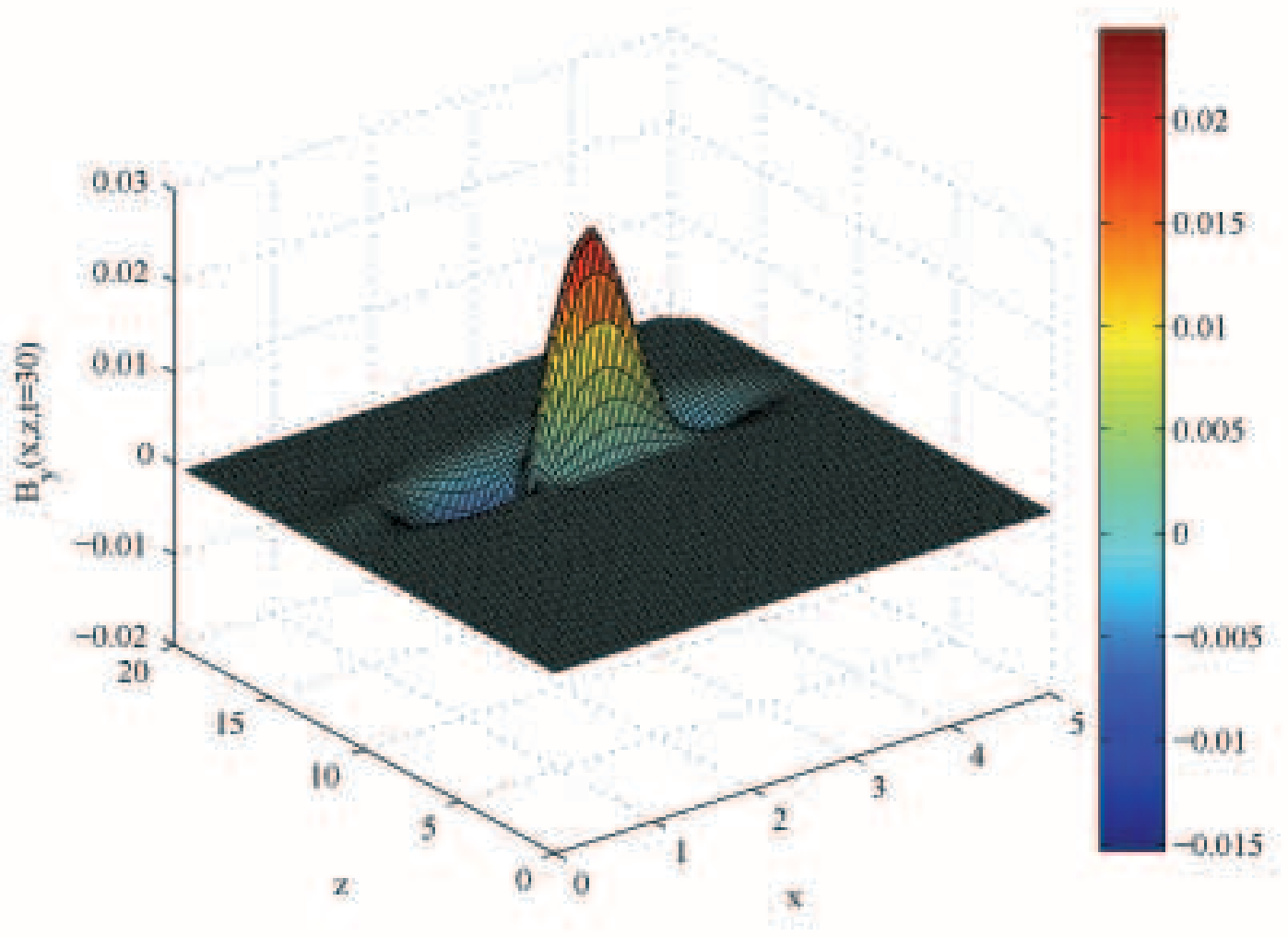}
\includegraphics[width=8cm]{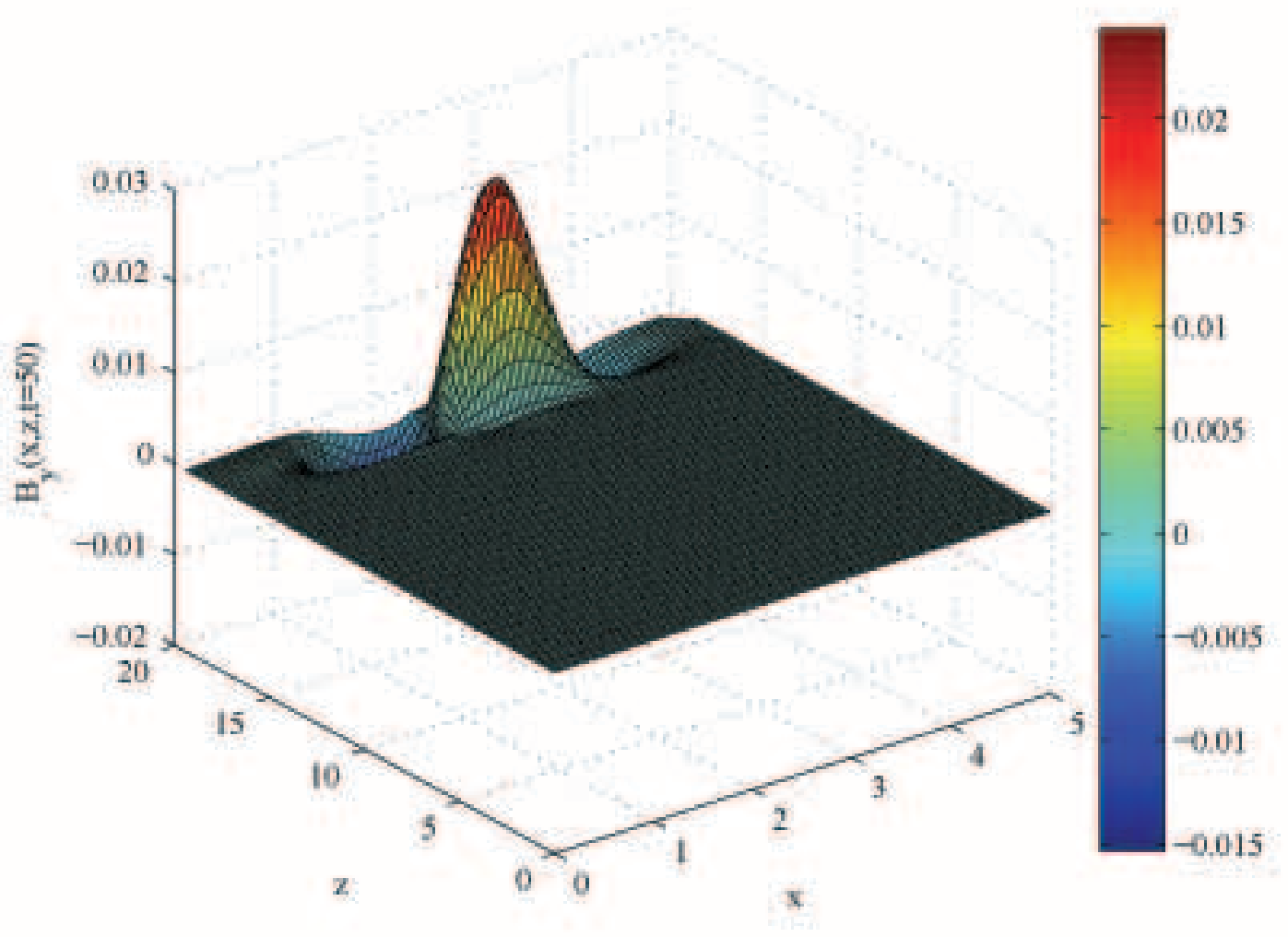}
\caption{The same as in Fig.~\ref{fig3} for the perturbed magnetic field. \label{fig4}}
\end{figure}

Figures~\ref{fig3} and \ref{fig4} illustrate the $3D$ plots of the perturbed velocity and magnetic field with respect to $x$, $z$ for $t= 5 \tau$~s,
$t= 30 \tau$~s, and $t= 50 \tau$~s. At the presence of the mentioned gradients and stratification due to gravity, the damping process takes place in time than in space.
In spite of the standing waves, propagating waves are stable and dissipate after some periods due to phase mixing \citep{Ebadi2012b}.


\begin{figure}
\centering
\includegraphics[width=8cm]{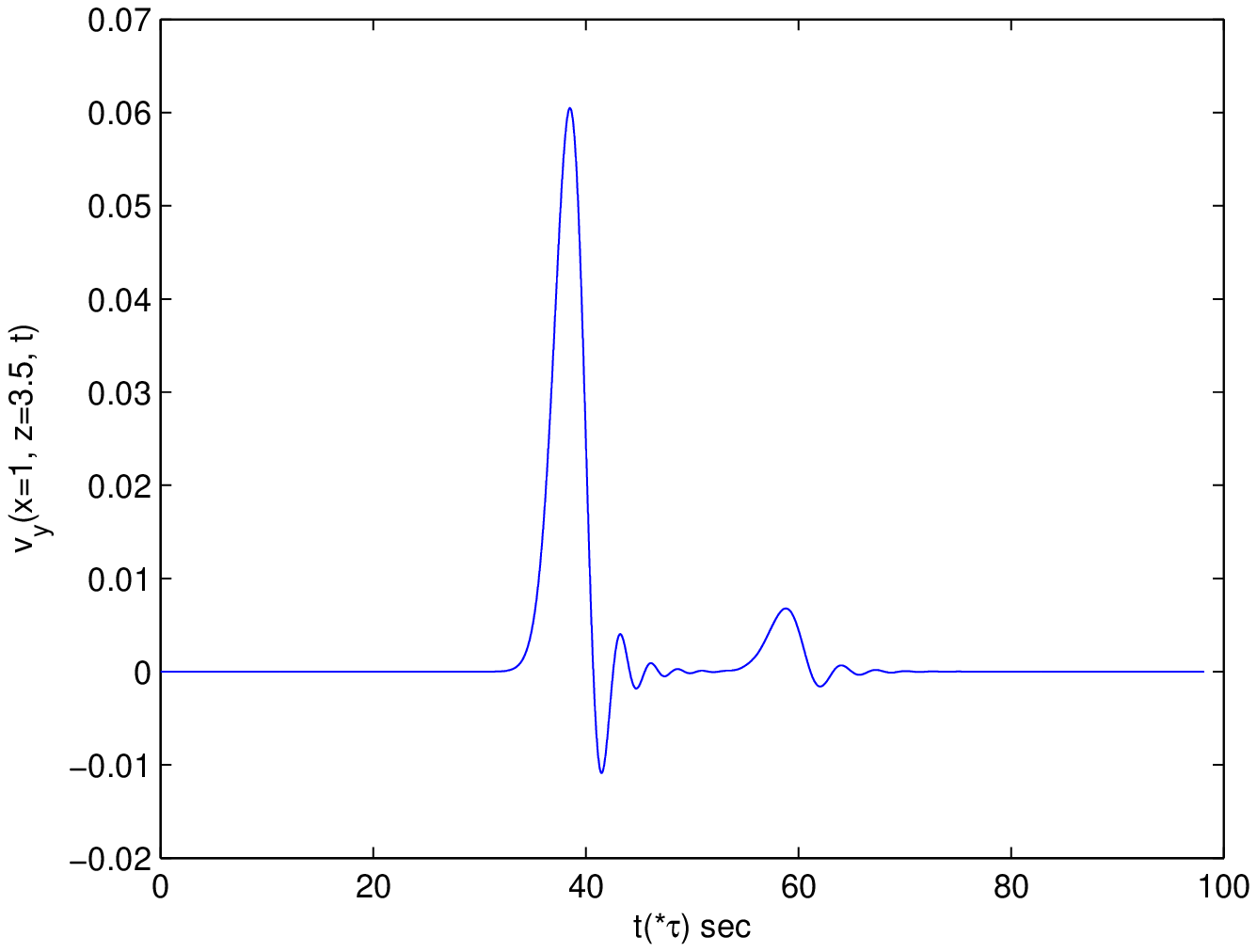}
\includegraphics[width=8cm]{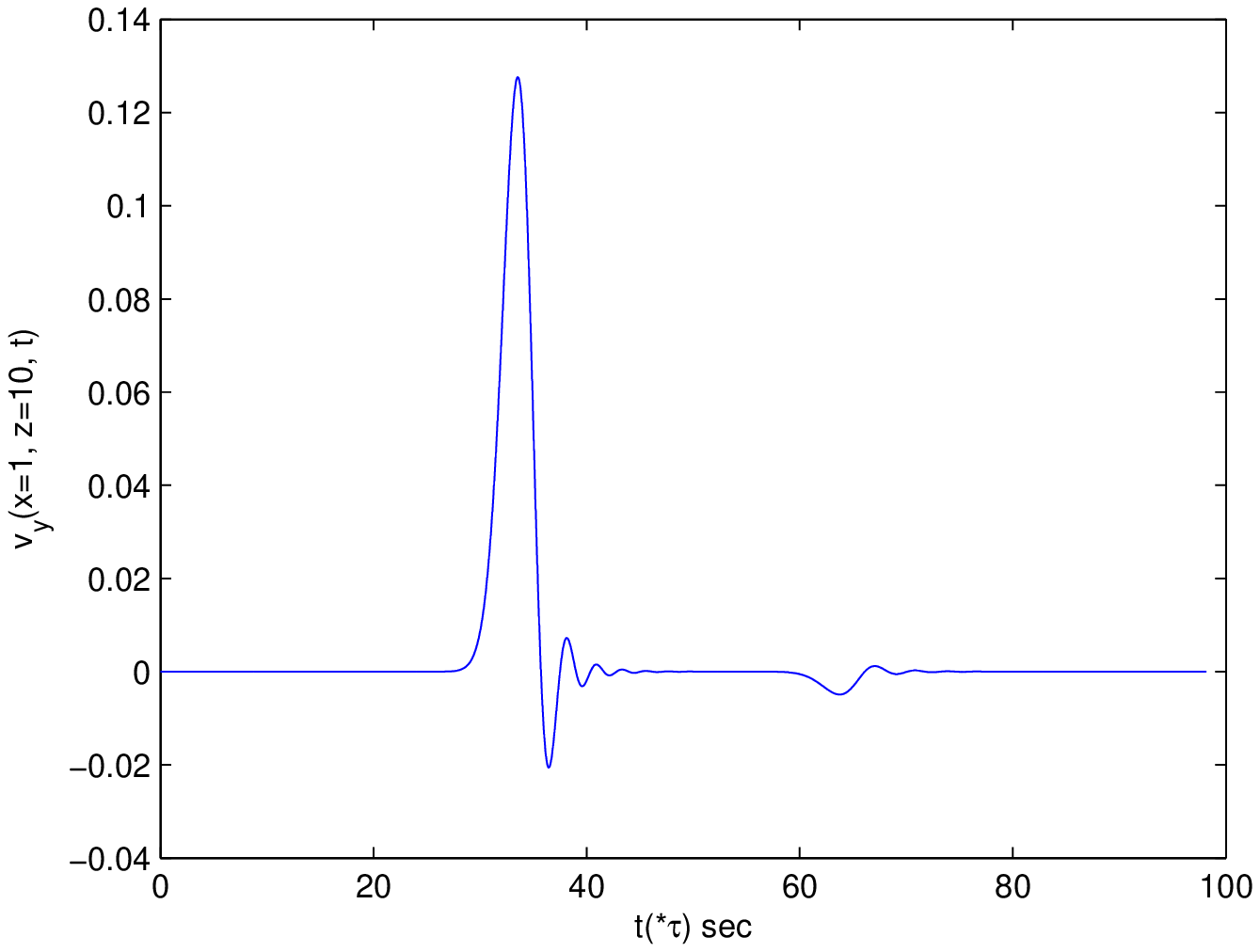}
\includegraphics[width=8cm]{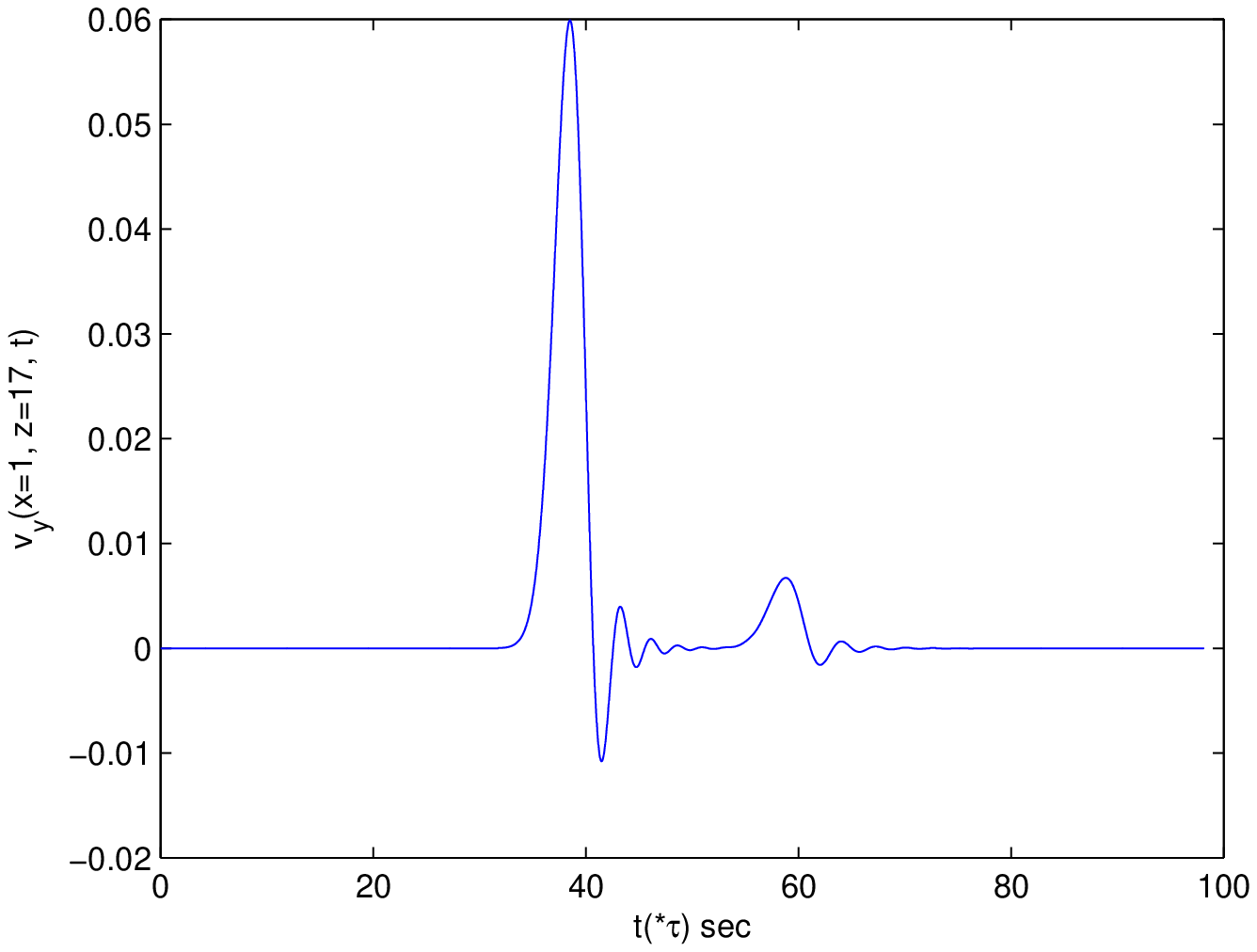}
\caption{The perturbed velocity variations are showed with respect to time and $x= 250$~km for three values of $z= 875$~km, $z= 2500$~km,
and $z= 4250$~km from top to bottom. \label{fig5}}
\end{figure}

\begin{figure}
\centering
\includegraphics[width=8cm]{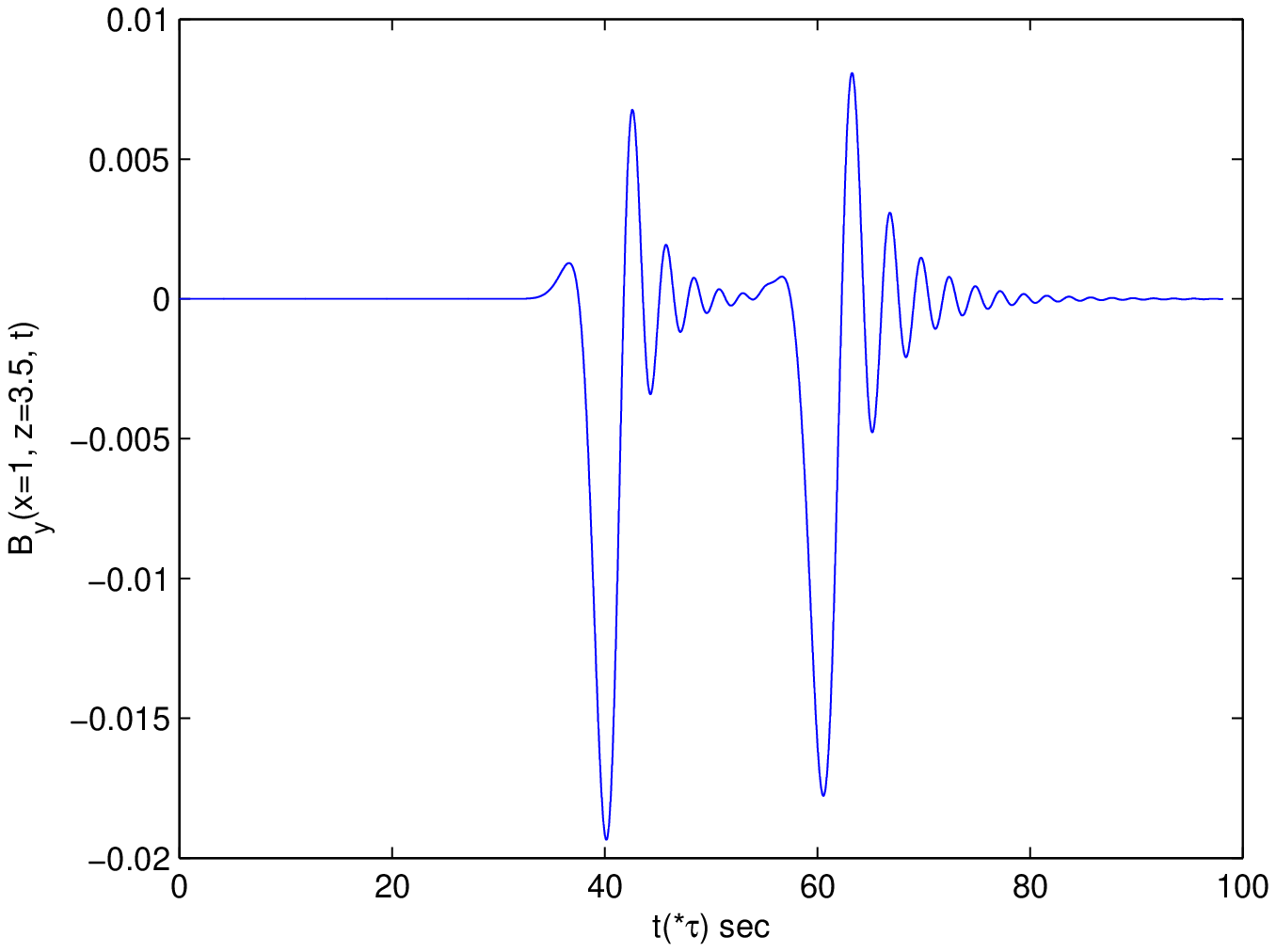}
\includegraphics[width=8cm]{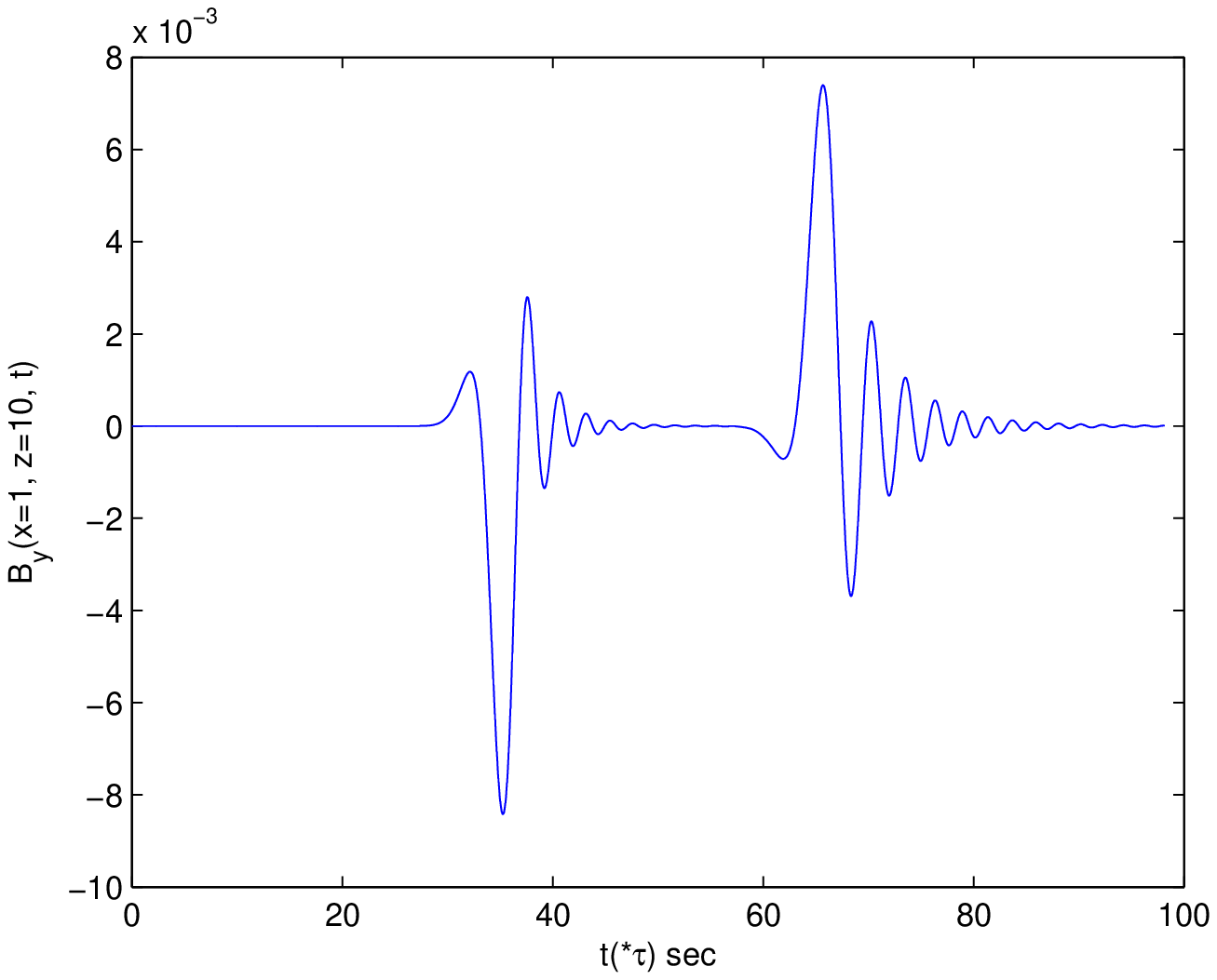}
\includegraphics[width=8cm]{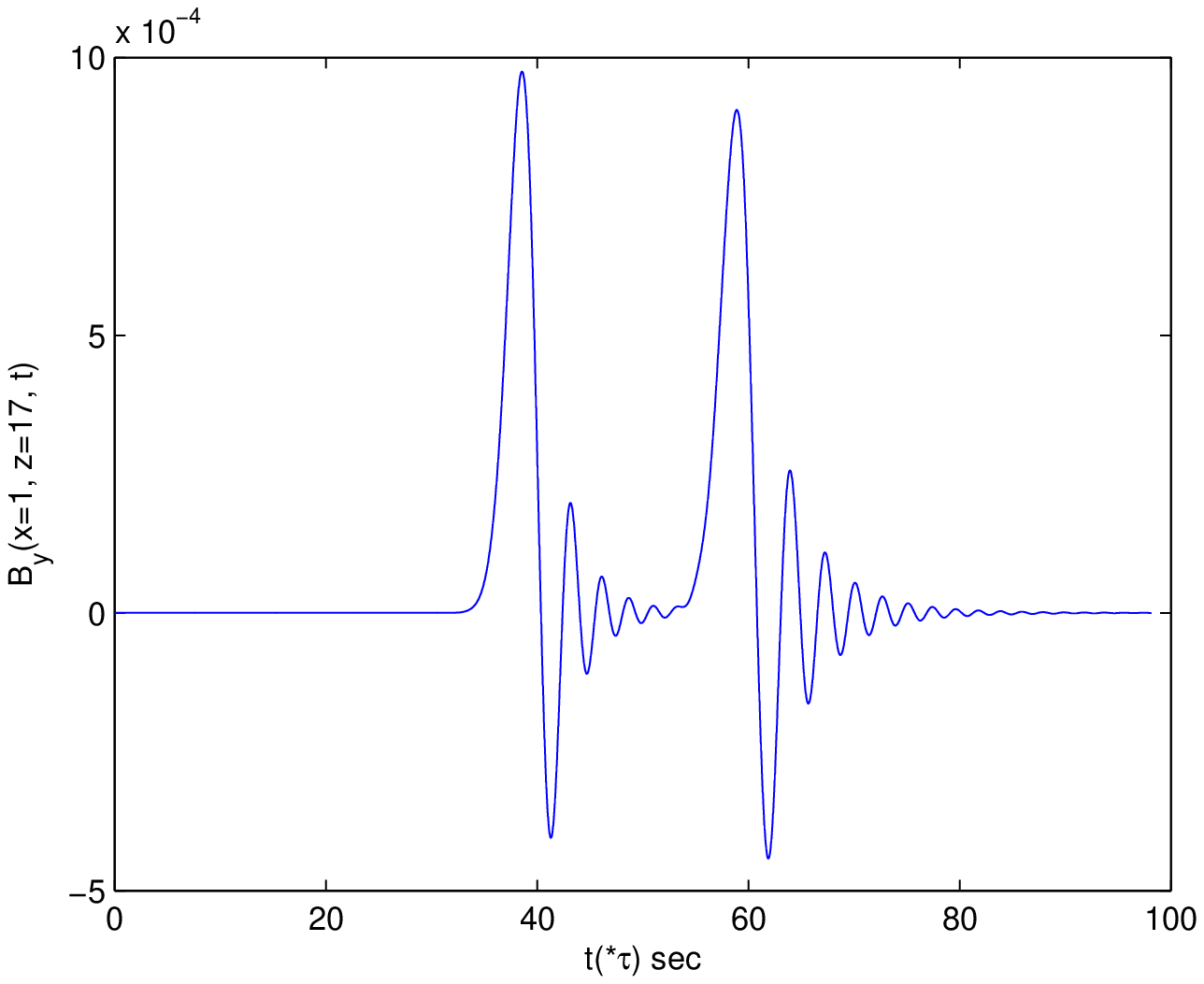}
\caption{The perturbed magnetic field variations are showed with the same coordinates as inferred in figure~\ref{fig5}. \label{fig6}}
\end{figure}

Figure~\ref{fig5} shows perturbed velocity variations with respect to time in $x=250$~km, $z=875$~km; $x=250$~km, $z=2500$~km;
and $x=250$~km, $z=4250$~km, respectively. In Figure~\ref{fig6}, perturbed magnetic field variations are presented for $x=250$~km,
$z=875$~km; $x=250$~km, $z=2500$~km; and $x=250$~km, $z=4250$~km, respectively. In these figures the perturbed velocity and magnetic
field are normalized to $V_{A0}$ and $B_{0}$ respectively, and it is obvious from the plots that there is a damping at the first stage of phase mixing.
This behavior can be related to the presence of transition region between chromosphere and corona.
\\At the first height ($z=875$~km), total amplitude of both velocity and magnetic field oscillations have values near to the initial ones.
As height increases, the perturbed velocity and magnetic field amplitudes increase. Nonetheless,
exponentially damping behavior is obvious in both cases. This means that with an increase in height, amplitude of velocity oscillations
is expanded due to significant decrease in density, which acts as inertia against oscillations. Similar results are observed by time-distance
analysis of solar spicule oscillations \citep{Ebadi2012a}. It is worth to note that the density stratification influence on the magnetic field
is negligible, which is in agreement with Solar Optical Telescope observations of solar spicules \citep{Verth2011}.


\begin{figure}
\centering
\includegraphics[width=8cm]{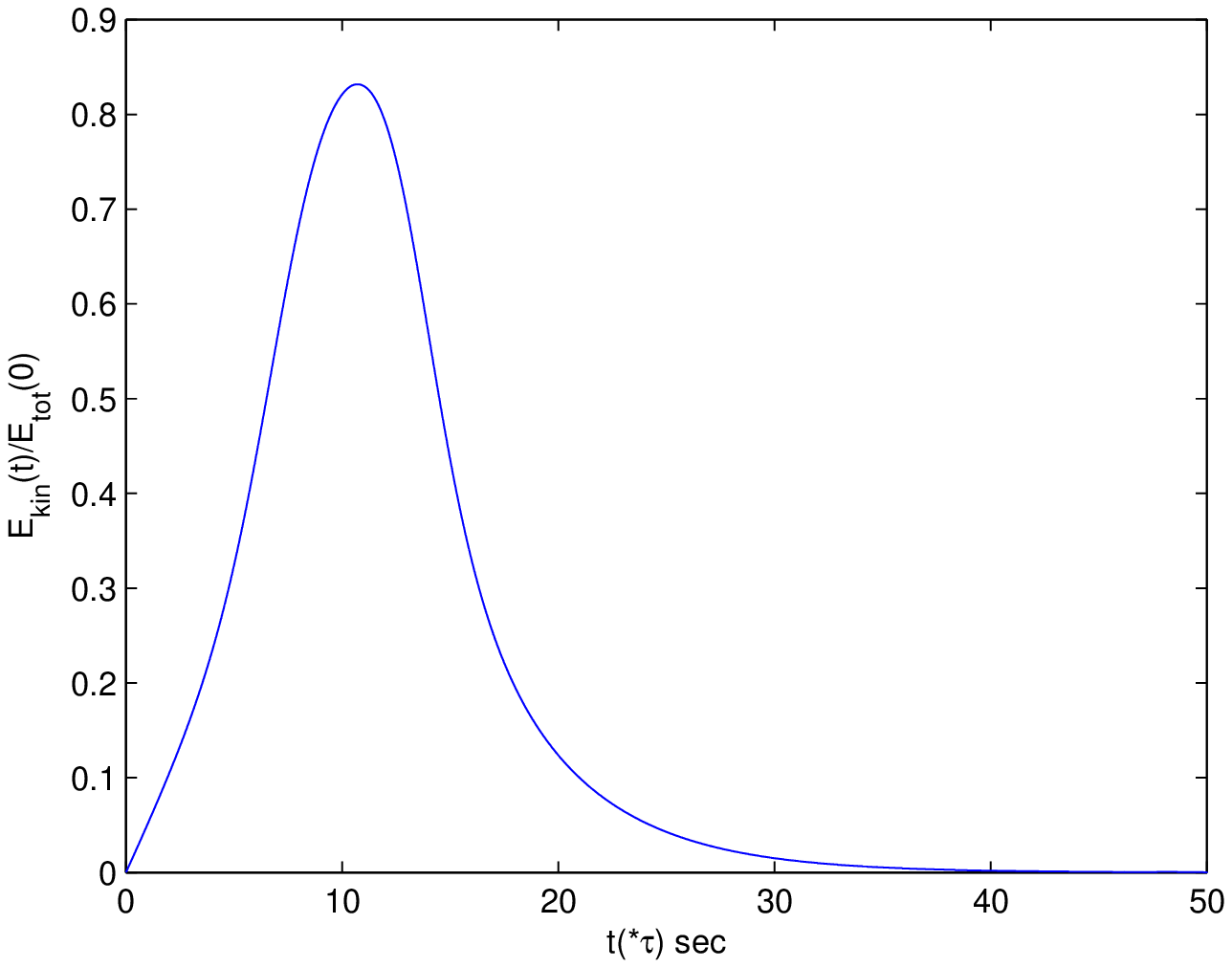}
\includegraphics[width=8cm]{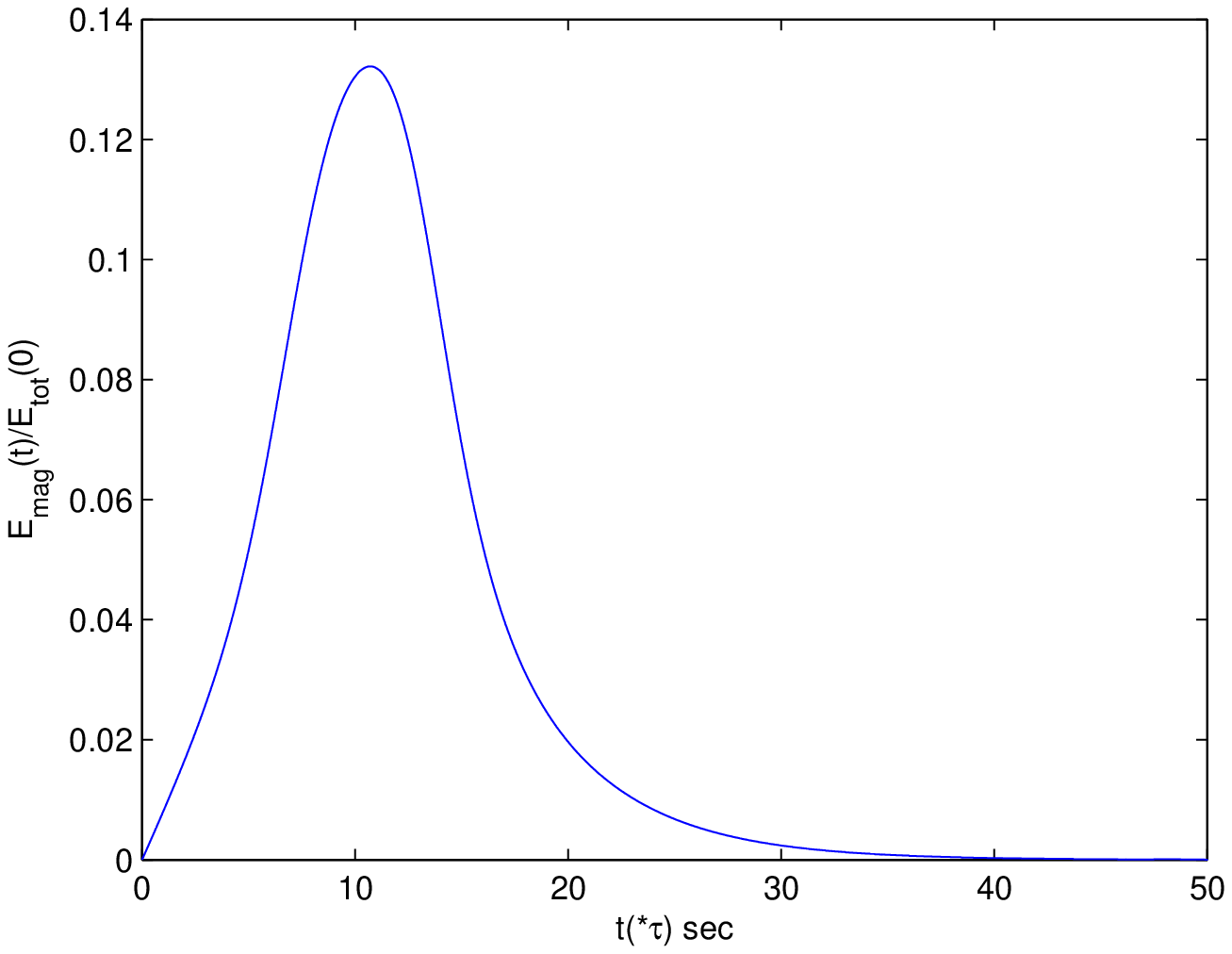}
\includegraphics[width=8cm]{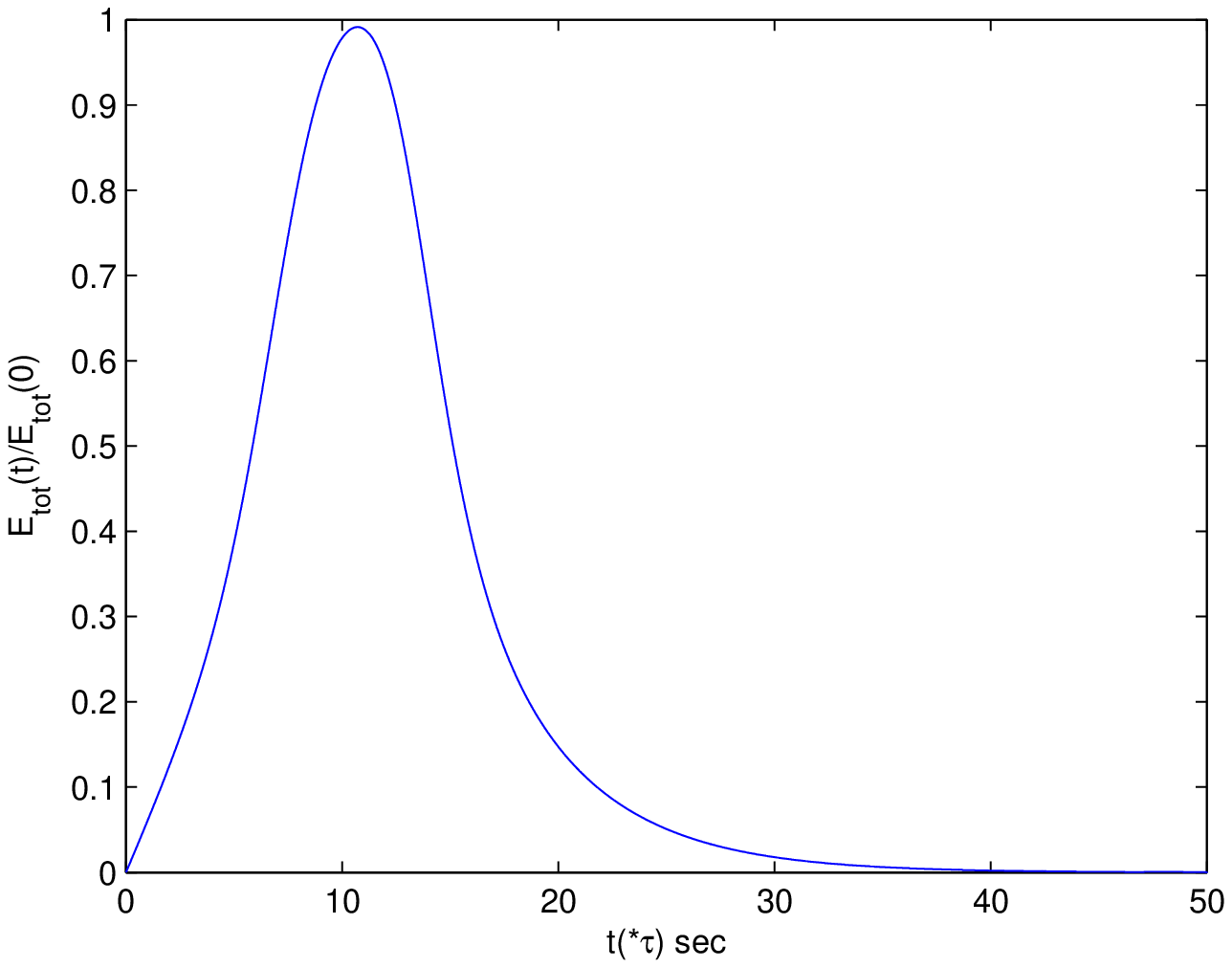}
\caption{Time variations of the normalized kinetic, magnetic and total energies for $\omega = 0.8a$. \label{fig7}}
\end{figure}

In Figure~\ref{fig7}, kinetic, magnetic and total energies normalized to the initial total energy, are presented from top to bottom, respectively.
Since the presence of transition region leads to the variation of viscosity and resistivity coefficients, these inhomogeneities can change the rate of damping
of Alfv\'{e}n waves. Thus, it seems that our model will be useful to study the solar spicules.
Spicules are short-living and transient phenomena, and we conclude that the phase mixing in such circumstances can occur in time rather than in space \citep{De99}.
Obtained damping times ($\tau_{d}= 320 s$) are in agreement with spicule lifetimes \citep{Tem2009}.

\citet{Ebadi2012b} studied the phase mixing in solar spicules with zero steady flows and vertical magnetic field.
At the first case they assumed non-ideal and linearized MHD equations with constant $v_{0}$ and $B_{0}$ in z direction only.
Then for comparison they run their simulations for the case $v_{0} =0$ and $B_{0z}$. They cannot found any significant difference
between these two cases which is discussed in the mentioned paper.
At the presence of transition region between chromosphere and corona and non-linear MHD equations, the damping times are shortened significantly.
This is very important because spicules disappear after a few periods, and the efficient damping treatment should result damping times as long as spicule lifetimes.
Moreover, the exponential behavior of energies is altered in non-linear case.

\subsection{Convergence of our numerical scheme}
In this section we check the convergence of our numerical simulation and the realistic behavior of the functions.
For the sake of simplicity we assume $\textbf{B} = B_{0}\hat{k}$ and $V_{0}=0$. The governed equations by these assumptions
are as follows:
\begin{equation}
\label{eq:mag1}
  \frac{\partial b}{\partial t} = \frac{\partial v}{\partial z} + \eta(z)(\frac{\partial^{2}}{\partial x^{2}}+\frac{\partial^{2}}{\partial z^{2}})b,
\end{equation}
and
\begin{equation}
\label{eq:vel1}
   \frac{\partial v}{\partial t} = v_{A}^2(x,z)\frac{\partial b}{\partial z} + \nu(z)(\frac{\partial^{2}}{\partial x^{2}}+\frac{\partial^{2}}{\partial z^{2}})v
\end{equation}
where $b$ and $v$ are y-components of the perturbed magnetic field and velocity, respectively.
Other quantities have the same definitions which are used in the text.
By using the finite diffidence method we arrive at these equations:
\begin{eqnarray}
\label{eq:veloci}
 v_{i,j,k+1} &=& \nonumber
 \\ v_{i,j,k} +\frac{v_{A_{i,j}}^2}{\Delta z}(b_{i,j+1,k}-b_{i,j,k}) \nonumber
 \\ + \frac{\nu \Delta t}{\Delta x^{2}}(v_{i+1,j,k}-2v_{i,j,k}+v_{i-1,j,k}) \nonumber
 \\ +\frac{\nu \Delta t}{\Delta z^{2}}(v_{i,j+1,k}-2v_{i,j,k}+v_{i,j-1,k}),
\end{eqnarray}
and
\begin{eqnarray}
\label{eq:magno}
 b_{i,j,k} &=& \nonumber
 \\b_{i,j,k-1}+\frac{\Delta t}{\Delta z}(v_{i,j+1,k-1}-v_{i,j,k-1}) \nonumber
 \\+\frac{\eta \Delta t}{\Delta x^{2}}(b_{i+1,j,k-1}-2b_{i,j,k-1}+b_{i-1,j,k-1}) \nonumber
 \\+\frac{\eta \Delta t}{\Delta z^{2}}(b_{i,j+1,k-1}-2b_{i,j,k-1}+b_{i,j-1,k-1})
\end{eqnarray}
where indices $i$,$j$, and $k$ corresponded to $x$, $z$, and $t$, respectively. $\Delta x$, $\Delta z$, and $\Delta t$ are
increments of $x$, $z$, and $t$, respectively.

According to \citet{Vasil1995}, $\nu \Delta t/\Delta z^{2}$ and $\eta \Delta t/\Delta z^{2}$ should be
smaller than $0.5$ for the convergence of simulations. It is obvious from our selected parameters in section $3$
that simulations are converged.

\section{Conclusion}
\label{sec:concl}
Study of Alfv\'{e}n waves in solar spicules may represent an efficient heating mechanism in the solar corona. In this paper we consider
spicules with steady flow and sheared magnetic field. The non-linear behavior of Alfv\'{e}n waves and their phase mixing has been studied at the presence of viscosity and resistivity gradients. These gradients are due to the presence of transition region between chromosphere and corona. Our numerical simulations show that the damping is enhanced related to viscosity and resistivity gradients. Moreover, the perturbed velocity and magnetic field amplitudes decrease with height and time in our non-linear model.
It is found that the non-linearity of the MHD equations changes the exponential behavior of energies.

\acknowledgments
This work has been supported financially by the Research Institute for Astronomy and
Astrophysics of Maragha (RIAAM), Maragha, Iran.

\makeatletter
\let\clear@thebibliography@page=\relax
\makeatother

\end{document}